\documentclass[preprint,amsmath,amssymb,aps,pra,longbibliography]{revtex4-1}

\usepackage[margin=0.75in, top=1in, bottom=0.85in]{geometry}

\usepackage[final,letterspace=-0]{microtype}
\usepackage[colorlinks=true, linkcolor=blue, urlcolor=blue, citecolor=blue, anchorcolor=blue]{hyperref}

\usepackage{mathpazo} % add possibly `sc` and `osf` options
\usepackage{upgreek}
\usepackage{xcolor}

\usepackage[]{graphicx}% Include figure files
\usepackage{dcolumn}% Align table columns on decimal point
\usepackage{bm}% bold math
\usepackage{helvet}

\begin{document}

\title{Free-space optical delay line using space-time wave packets}

\author{Murat Yessenov, Basanta Bhaduri, Peter J. Delfyett, and Ayman F. Abouraddy}

\affiliation{CREOL, The College of Optics \& Photonics, University of Central Florida, Orlando, FL 32816, USA}

\begin{abstract}
An optical buffer having a large delay-bandwidth-product -- a critical component for future all-optical communications networks -- remains elusive. Central to its realization is a controllable inline optical delay line, previously accomplished via engineered dispersion in optical materials or photonic structures constrained by a low delay-bandwidth product. Here we show that space-time wave packets whose group velocity in free space is continuously tunable provide a versatile platform for constructing inline optical delay lines. By spatio-temporal spectral-phase-modulation, wave packets in the same or in different spectral windows that initially overlap in space and time subsequently separate by multiple pulse widths upon free propagation by virtue of their different group velocities. Delay-bandwidth products of $\sim100$ for pulses of width $\sim1$~ps are observed, with no fundamental limit on the system bandwidth.
\end{abstract}

\maketitle

%\section{Introduction}

The relentless increase in demand for communications bandwidth \cite{Richarsdon10Sc} has spurred developments in sub-systems that are critical for future optical networks, such as spatial-mode multiplexing \cite{Bozinovic13Science,Li14AOP,Zhao15NP} and ultrafast modulators \cite{Wang18Nature}. A critical -- but to date elusive -- component is an optical buffer that alleviates data packet contention at optical switches by reordering the packets without an optical-to-electronic conversion  \cite{Boyd06OPN}. Previous efforts addressing this challenge have exploited so-called `slow light' \cite{Krauss08NP,Boyd09Science}, which refers to the reduction in the group velocity of a pulse traversing a selected material \cite{Hau99Nature,Kash99PRL,Song05OE,Okawachi05PRL} or carefully designed photonic structure \cite{Vlasov05N,Krauss07JPD,Baba08NP}. Despite the diversity of their physical embodiments \cite{Kurgin08Book}, slow-light approaches typically rely on resonant optical effects and are thus limited by a delay-bandwidth product (DBP) on the order of unity. That is, the differential group delay with respect to a reference pulse traveling at $c$ (the speed of light in vacuum) does not exceed the pulse width \cite{Tucker05EL,Tucker05JLT}, which falls short of the requirements of an optical buffer \cite{Parra07OPN}. Time-varying systems \cite{Yanik04PRL} can overcome this limit at the expense of increased implementation complexity. In one realization, a trapdoor mechanism in coupled resonators increases the storage time through externally controlled coupling \cite{Xu06NPhys} -- but losses concomitantly increase in step with the delay. An alternate approach makes use of non-resonant recycling of orthogonal modes in a cavity \cite{Chang17arxiv}. A different strategy relies on transverse spatial structuring to reduce the group velocity in free space, but only a minute reduction below $c$ has been detected to date \cite{Giovannini15Science,Bouchard16Optica}. Nevertheless, theoretical proposals suggest that pushing this approach to the limit may produce sufficiently large differential group delays for an optical buffer \cite{Zapata08JOSAA,Alfano16}, but temporal spreading is associated with the propagation of these wave packets \cite{Saari17OC}. Finally, a recent theoretical proposal suggests that optical non-reciprocity can help bypass the usual DBP limits \cite{Tsakmakidis17Science2}, but doubts have been cast on this prospect \cite{Tsang18OL,Mann19Optica}.

An altogether different approach for tuning the group velocity of a pulse makes use of 'space-time' (ST) wave packets: propagation-invariant pulsed beams (diffraction-free and dispersion-free) \cite{Brittingham83JAP,Ziolkowski85JMP,Lu92IEEEa,Saari97PRL,Reivelt03arxiv,Kiselev07OS,Turunen10PO,FigueroaBook14} endowed with structured spatio-temporal spectra \cite{Donnelly93PRSLA,Porras17OL,Efremidis17OL,Kondakci17NP,PorrasPRA18} in which each spatial frequency is associated with a single wavelength \cite{Longhi04OE,Saari04PRE,Kondakci16OE,Parker16OE,}. Although it has long been known theoretically that ST wave packets may take on arbitrary group velocities (speed of the wave-packet peak) in free space \cite{Salo01JOA,Zapata06OL,Valtna07OC,Zamboni08PRA,Wong17ACSP2}, experiments have revealed group-velocity deviations from $c$ of only $\sim0.1\%$ \cite{Bonaretti09OE,Bowlan09OL,Kuntz09PRA}, corresponding to group delays of tens or hundreds of femtoseconds -- several orders-of-magnitude below the requirements for an optical buffer. We recently introduced a spatio-temporal phase-only synthesis strategy for the preparation of ST wave packets that allows fully exploiting their unique characteristics \cite{Yessenov19OPN}. For example, we have reported free-space group velocities tunable from $30c$ to $-4c$ \cite{Kondakci19NC}, group delays of $\sim150$~ps \cite{Yessenov19OE}, propagation distances extending to 70~m \cite{Bhaduri18OE,Bhaduri18OL}, among other unique properties \cite{Kondakci18PRL,Kondakci18OL,Bhaduri19Optica,Kondakci19ACSP}. These attributes indicate the potential utility of ST wave packets in constructing an optical buffer.

Here we report a proof-of-principle realization of a readily controllable, inline, free-space optical delay line -- the centerpiece of an optical buffer -- utilizing the tunable group velocity of ST wave packets. By assigning to optical pulses different group velocities (whether subluminal, superluminal, or negative), the resulting wave packets separate from each other upon free propagation by multiple pulse widths. We observe a differential group delay of $\sim\!160$~ps between two wave packets of widths as small as $\sim\!1.3$~ps, corresponding to a DBP of $\sim\!100$, which is $\sim\!2$~orders-of-magnitude larger than previous demonstrations \cite{Parra07OPN}. We realize several desiderata for an all-optical buffer, such as introducing a controllable delay between two pulses (or more) in the same or different spectral channels, while occupying the same or different spatial modes. There is no need here for the widths of the spectral channels to be equal nor do they need to be mutually coherent. It is critical to note that tuning the group velocity is \textit{not} associated with changes in the pulse amplitude. Furthermore, subluminal and superluminal wave packets are treated on the same footing, in contrast to the typically distinct approaches of slow-light and fast-light, which highlights the versatility of our strategy. Crucially, the group delay introduced between the wave packets in free space is \textit{independent} of the external degrees of freedom of the field (such as wavelength). Moreover, due to the absence of limitations stemming from resonant effects, there are in principle no bandwidth constraints. Although the spectral resolution of the system imposes a limit on the maximal differential group delay \cite{Yessenov19OE}, there are no limits on the pulse width that can be exploited. Consequently, future increases in transmission rates can be accommodated without changing the system architecture.

\section*{Results}

\subsection*{Concept of a free-space optical delay line based on ST wave packets}

Our overall conceptual scheme is illustrated in Fig.~\ref{Fig:Concept}a. Optical pulses that overlap in time and travel in free space at a group velocity $\widetilde{v}\!=\!c$ are transformed into ST wave packets having different group velocities $\widetilde{v}_{1}$ and $\widetilde{v}_{2}$ via spatio-temporal spectral phase modulations implemented independently in each spectral channel. By virtue of their different group velocities, the wave packets gradually separate in time by a group delay that depends on the difference between their group velocities and the propagation distance.

The impact of the phase modulation can be appreciated by examining the spectral space $(k_{x},k_{z},\tfrac{\omega}{c})$, where $k_{x}$ and $k_{z}$ are the transverse and axial components of the wave vector, and $\omega$ is the angular frequency, which satisfy the free-space dispersion relationship $k_{x}^{2}+k_{z}^{2}\!=\!(\tfrac{\omega}{c})^{2}$ that is represented geometrically as the light-cone. We assume the field is uniform along the other transverse coordinate $y$ ($k_{y}\!=\!0$), and refer to $k_{x}$ and $\omega$ as the spatial and temporal frequencies, respectively. The spatio-temporal spectra of the initial plane-wave pulses lie along the tangent to the light-cone at $k_{x}\!=\!0$. The spectral phase modulation is designed to rotate the tangential plane by a spectral tilt angle $\theta$ with respect to the $k_{z}$-axis. The result is a propagation-invariant wave packet traveling at a group velocity of $\widetilde{v}\!=\!c\tan{\theta}$, where $\widetilde{v}$ is the velocity of the central peak of the wave packet. Although $\widetilde{v}$ can take on arbitrary values, this does \textit{not} entail a violation of relativistic causality \cite{Shaarawi00JPA,SaariPRA18,Saari19PRA,Yessenov19OE}, as we explain further below. We delineate three regimes \cite{Yessenov19PRA}: $0\!<\!\theta\!<\!45^{\circ}$ is subluminal $\widetilde{v}\!<\!c$; $45^{\circ}\!<\!\theta\!<\!90^{\circ}$ is superluminal $\widetilde{v}\!>\!c$; and $90^{\circ}\!<\!\theta\!<\!180^{\circ}$ is a negative-$\widetilde{v}$ domain $\widetilde{v}\!<\!0$. In the subluminal regime, the spatio-temporal spectral locus of the wave packet is an ellipse, and is a hyperbola in the superluminal and negative-$\widetilde{v}$ regimes (Fig.~\ref{Fig:Concept}a, insets); see Methods.

Previous experiments \cite{Saari97PRL,Reivelt02PRE,Faccio07OE}, including our group's \cite{Kondakci17NP,Kondakci19NC,Bhaduri19Optica,Yessenov19OE}, have all been limited to synthesizing single ST wave packets of a given group velocity. Here we synthesize up to three ST wave packets inline simultaneously -- each wave packet having a \textit{different} group velocity -- by implementing a segmented phase modulation scheme that addresses each spectral channel separately (Fig.~\ref{Fig:Concept}a, inset). Therefore, the initially overlapping wave packets travel rigidly in free space at their respective group velocities without diffraction or dispersion, thereby separating from each other (Methods). A large number of wave packets can be addressed in this fashion using large-area phase plates for spectral-phase modulation \cite{Kondakci18OE}, and the widths of the spectral channels need not be equal nor the spectra have mutual coherence \cite{Yessenov19Optica,Yessenov19OL}.

The propagation dynamics of ST wave packets having different group velocities is depicted in Fig.~\ref{Fig:Concept}b,c. The wave packets overlap in time at $z\!=\!0$, but no longer overlap at later times: the superluminal wave packet has moved the farthest ahead in space (beyond the distance traveled by a luminal pulse); the subluminal  wave packet lags behind; while a negative-$\widetilde{v}$ wave packet is located at a negative-$z$ position (Fig.~\ref{Fig:Concept}b). In our experiments, rather than detecting the ST wave packets at different $z$ for fixed delay, we instead scan the arrival time $\tau$ at a fixed detector position (Fig.~\ref{Fig:Concept}c). By moving the detector to a positive-$z$ location, the superluminal wave packet arrives earlier than a luminal pulse while the subluminal wave packet arrives later. The negative-$\widetilde{v}$ wave packet arrives earlier at $z\!>\!0$ than at $z\!=\!0$, thereby indicating its backward motion -- as confirmed unambiguously in our measurements.

\subsection*{Experimental arrangement}

Our experiments (setup shown in Fig.~\ref{Fig:Concept}d) are carried out using plane-wave femtosecond pulses at a wavelength of $\sim\!800$~nm from a Ti:sapphire laser (Tsunami, Spectra Physics). Spatio-temporal spectral modulation is performed with a diffraction grating (Newport 10HG1200-800-1) and a spatial light modulator (SLM; Hamamatsu X10468-02) \cite{Kondakci17NP,Kondakci19NC,Bhaduri19Optica,Yessenov19OPN}. A bandwidth of $\sim1.5$~nm is spread over the width of the SLM active area, which is subdivided into three spectral channels (Ch1, Ch2, and Ch3) each of width $\sim0.5$~nm, each channel therefore corresponding to pulses of width $\sim 4$~ps. The measured spectra of the three channels are plotted in Fig.~\ref{Fig:Spectra}a. In contrast to the one-dimensional (1D) phase distributions exploited in traditional ultrafast pulse modulation \cite{Weiner00RSI,Weiner09Book}, here a two-dimensional (2D) phase distribution jointly manipulates the spatial \textit{and} temporal spectra. Within each spectral channel the 2D phase distribution produces a ST wave packet having a spectral tilt angle of $\theta$, as shown in Fig.~\ref{Fig:Spectra}b. We assign Ch1 the spectral tilt angle $\theta_{1}\!=\!30^{\circ}$, corresponding to the subluminal group velocity $\widetilde{v}_{1}\!=\!0.58c$; Ch2 is superluminal with $\theta_{2}\!=\!70^{\circ}$ and $\widetilde{v}_{2}\!=\!2.75c$; whereas Ch3 is negative-$\widetilde{v}$ with $\theta_{3}\!=\!120^{\circ}$ and $\widetilde{v}\!=\!-1.73c$. Any other values could be implemented in these channels by simply adjusting the phase distribution, and any channel can be deactivated by setting the SLM phase distribution to zero. A spatial filter in the path of the wave packets eliminates any deactivated channel.

The spatio-temporal spectral intensity of the ST wave packet is recorded in $(k_{x},\lambda)$ space by implementing a spatial Fourier transform on the spectrally resolved pulses \cite{Kondakci17NP}. We plot in Fig.~\ref{Fig:Spectra}c the spatio-temporal spectra of the three channels captured simultaneously. The \textit{curvature} of each spectrum is determined by the associated spectral tilt angle $\theta$ \cite{Faccio07OE}. From this spectral measurement we obtain the spectral projection onto the $(k_{z},\tfrac{\omega}{c})$ space; see Fig.~\ref{Fig:Spectra}d. The slopes of the linear spectral projections onto the $(k_{z},\tfrac{\omega}{c})$ space provide independent confirmation of the group velocities ($\widetilde{v}\!=\!\tfrac{\partial\omega}{\partial k_{z}}$). Finally, by recording the time-averaged intensity distribution of the ST wave packet upon propagation $I(x,z)\!=\!\int\!dt|E(x,z,t)|^{2}$ with a CCD camera (The ImagingSource, DMK 33UX178), we confirm the diffraction-free propagation as shown in Fig.~\ref{Fig:Spectra}e for Ch2 (Ch1 and Ch3 are deactivated). We define $L_{\mathrm{max}}$ to be the distance at which $I(0,L_{\mathrm{max}})\!=\!\tfrac{1}{e}I(0,0)$.

To characterize the spatio-temporal intensity profile of ST wave packets and measure their group velocity, we place the spatio-temporal synthesis arrangement in one arm of a Mach-Zehnder interferometer \cite{Kondakci19NC,Bhaduri19Optica} and place a temporal delay in the other arm in the path of a reference pulse (100-fs pulses from the Ti:sapphire laser); see Fig.~\ref{Fig:Concept}d. Such a measurement configuration corresponds to the conceptual scheme depicted in Fig.~\ref{Fig:Concept}c. This technique yields only the intensity of the wave packet profile and not its phase, which requires more sophisticated spatio-temporal characterization tools \cite{Bowlan09OL,Adams09OL,Lohmus12OL}. Nevertheless, our approach suffices for estimating the arrival times and thence the group velocities \cite{Kondakci19NC,Bhaduri19Optica}.

As seen in Fig.~\ref{Fig:Spectra}c (and also in Fig.~\ref{Fig:Spectra}d), the ideal correlation function between $k_{x}$ and $\omega$ is relaxed, mainly due to the finite spectral resolution of the diffraction gratings used. Ideal ST wave packets propagate indefinitely at an arbitrary group velocity but require infinite energy, whereas realistic finite-energy ST wave packets maintain such group velocities over a bounded distance. Finite-energy ST wave packets are characterized by a spectral uncertainty $\delta\omega$ ($\delta\lambda$ in wavelength), which corresponds to an unavoidable fuzziness in the association between spatial and temporal frequencies. In our experiments, we estimate a spectral uncertainty $\delta\lambda\!\sim\!25$~pm \cite{Yessenov19OE}. Within this conception, a finite-energy ST wave packet can be decomposed into a product of an ideal ST wave packet traveling indefinitely at $\widetilde{v}$ (with a pulse width at the beam center proportional to the inverse of the bandwidth $\Delta\omega$) and a broad `pilot envelope' propagating at a group velocity of $c$ (and whose width in time is the inverse of the spectral uncertainty $\delta\omega$). Temporal walk-off between the ideal ST wave packet and the pilot envelope limits the propagation distance $L_{\mathrm{max}}$ of the ST wave packet and hence also constrains the maximum differential group delay $\tau_{\mathrm{max}}$ achieved with respect to a reference pulse traveling at $c$. It can be shown that $\tau_{\mathrm{\max}}\!\sim\!\tfrac{1}{\delta\omega}$ and $L_{\mathrm{max}}\!\sim\!\tfrac{c\tau_{\mathrm{max}}}{|1-\cot{\theta}|}$ \cite{Yessenov19OE}. Therefore, two ST wave packets having the same spectral uncertainty will travel different distances $L_{\mathrm{max}}$ according to their respective spectral tilt angles. The pilot envelope traveling at $c$ ensures that no violation of relativistic causality occurs even for ST wave packets propagating at superluminal or negative $\widetilde{v}$. In such cases, when the wave packets reach the edge of the pilot envelope, they are suppressed and their spatio-temporal profile deformed \cite{Yessenov19OE}. This is associated with the drop in on-axis intensity values as shown in Fig.~\ref{Fig:Spectra}e. Figure~\ref{Fig:Concept}b,c illustrate the propagation dynamics of three ST wave packets along the pilot envelope.

\subsection*{Measurement results}

\begin{figure*}[t!]
  \begin{center}
  \includegraphics[width=17.6cm]{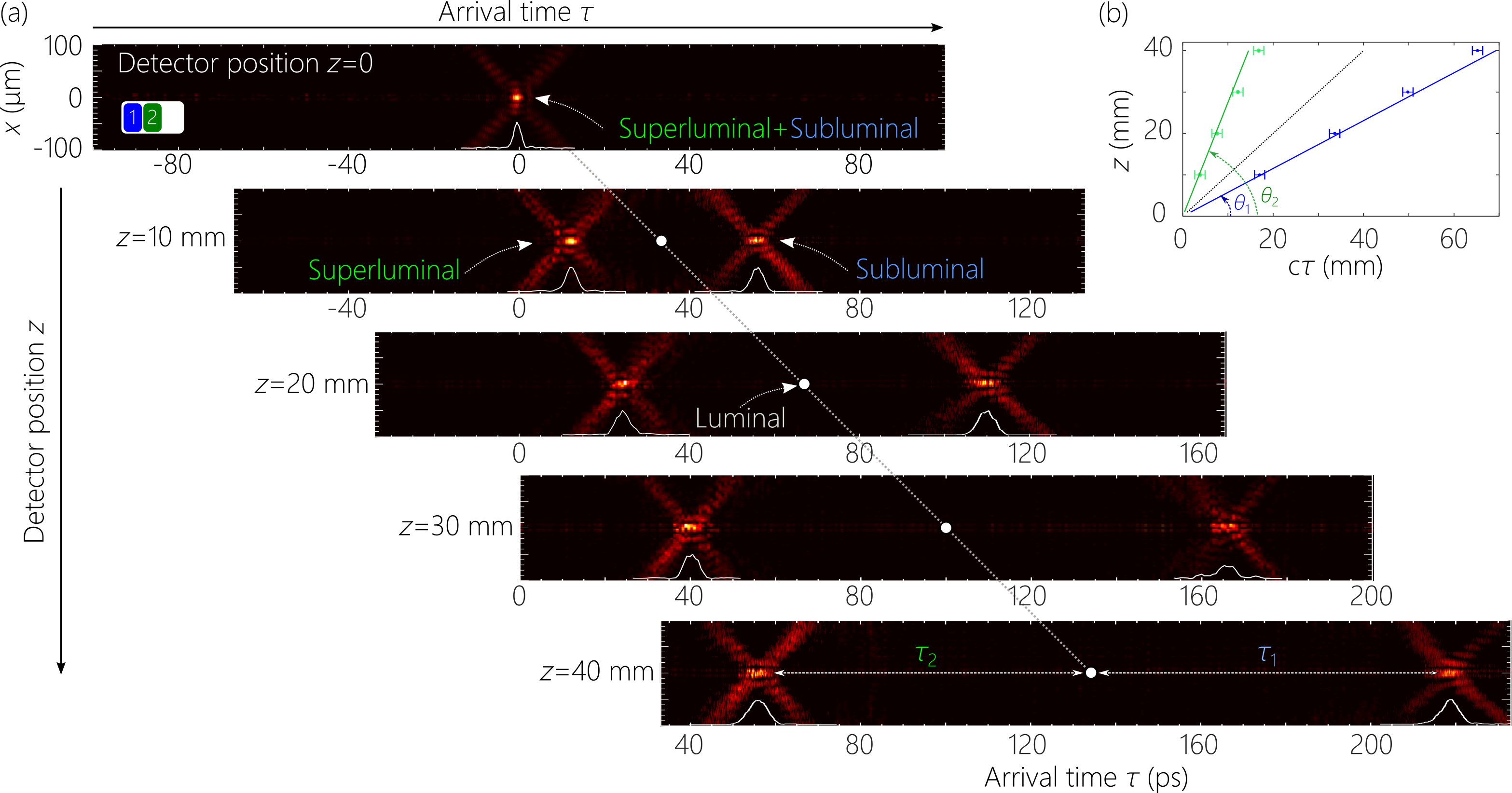}
  \end{center}
  \caption{Propagation of two ST wave packets having different group velocities in free space. (a) The panels display the arrival time of two wave packets (subluminal Ch1 with $\theta_{1}\!=\!30^{\circ}$ and superluminal Ch2 with $\theta_{2}\!=\!70^{\circ}$) at the detector plane $z$. When the detector is displaced to $z\!=\!0$, the two wave packets arrive simultaneously at $\tau\!=\!0$. When the detector is placed at $z\!=\!40$~mm, the superluminal wave packet arrives at $\tau\!\approx\! 185$~ps while the subluminal wave packet arrives at $\tau\!\approx\!23$~ps. In each panel we also plot the pulse profile at the wave packet center $x\!=\!0$, $I(0,z,\tau)$. The vertical and horizontal scales are adjusted so that the slope of the line traced by the center of each wave packet is $\tan{\theta}$, where $\theta$ is the spectral tilt angle. (b) Dynamics of the centers of the two wave packets revealing group velocities of $\widetilde{v}_{1}\!=\!c\tan{\theta_{1}}$ and $\widetilde{v}_{2}\!=\!c\tan{\theta_{2}}$. The points are experimental data and the straight lines are theoretical fits. The intermediate tilted line corresponds to a luminal pulse with $\widetilde{v}\!=\!c$.}
  \label{Fig:PulsePropagation2}
\end{figure*}

We start by discussing the results of monitoring the propagation of a pair of ST wave packets of different group velocities (subluminal Ch1 and superluminal Ch2) in free space using our interferometric approach, which are plotted in Fig.~\ref{Fig:PulsePropagation2}. The spatio-temporal intensity profile is traced by scanning a 200-ps delay line in the reference arm of the Mach-Zehnder interferometer (Fig.~\ref{Fig:Concept}d) with the detector placed at different locations along $z$. The delay $\tau$ in the path of the reference pulse is adjusted first so that the reference overlaps with the two ST wave packets when the detector is placed at $z\!=\!0$. At this point, the two ST wave packets overlap in space and time. Note the X-shaped spatio-temporal profile that is characteristic of most ST wave packets \cite{Saari97PRL}. After moving the detector to $z\!=\!10$~mm, the wave packets are found to arrive at different times, which is verified by scanning the reference delay $\tau$. The superluminal pulse (on the left) arrives early and the subluminal pulse (on the right) arrives late with respect to a pulse traveling at $c$ whose position is indicated by a white circle in Fig.~\ref{Fig:PulsePropagation2}a. The delay between the two pulse centers arriving at axial position $z$ is $\Delta\tau\!=\!\tfrac{z}{\tilde{v}_{1}}-\tfrac{z}{\tilde{v}_{2}}\!=\!\tfrac{z}{c}\{\cot{\theta_{1}}-\cot{\theta_{2}}\}$. Therefore we expect that $\Delta\tau\!=\!45.6$~ps when the detector is placed at $z\!=\!10$~mm and $\Delta\tau\!=\!182$~ps when $z\!=\!40$~mm; the corresponding measured values are $\Delta\tau\!\approx\!44$~ps and $\Delta\tau\!\approx\!162$~ps, respectively. We note that the group delays far exceed the temporal width $\Delta\tau$ of either wave packet at the center $x\!=\!0$, which corresponds to an effective delay-bandwidth product $\mathrm{DBP}\gg1$. The dynamics of the pulse centers are plotted in Fig.~\ref{Fig:PulsePropagation2}b showing clearly the difference in their group velocities with respect to $c$. This is the first observation of two co-propagating optical wave packets separating from each other in free space without the influence of an optical material or photonic structure. Note that the spectral uncertainties for both ST wave packets are equal because their values are set by the grating size.

We next examine the dynamics of a different configuration where Ch1 and Ch3 are activated while Ch2 is deactivated, corresponding to subluminal and negative-$\widetilde{v}$ wave packets. As seen in Fig.~\ref{Fig:PulsePropagation3}a, the subluminal wave packet propagates in the forward direction while the negative-$\widetilde{v}$ wave packet travels \textit{backwards} and arrives at the detector plane $z\!>\!0$ at $\tau\!<\!0$. Finally, we activate Ch1, Ch2, and Ch3 simultaneously and observe their evolution, which is plotted in Fig.~\ref{Fig:PulsePropagation3}b. The dynamics of the wave packet centers as plotted in the insets of Fig.~\ref{Fig:PulsePropagation3}a and Fig.~\ref{Fig:PulsePropagation3}b confirm that the targeted group velocities have been realized. Once again, the spectral uncertainty is the same for all three channels.

After investigating the dynamics of ST wave packets sharing the \textit{same} spatial mode but different spectral channels, we move on to the case of ST wave packets in the \textit{same} spectral channel. First, two ST wave packets sharing Ch1 are assigned different group velocities ($\widetilde{v}_{1}$ and $\widetilde{v}_{2}$) by \textit{interleaving} the spectral phase patterns corresponding to the targeted spectral tilt angles ($\theta_{1}$ and $\theta_{2}$); see Fig.~\ref{Fig:PulsePropagation3}c, inset. The measured spatio-temporal spectrum in $(k_{x},\lambda)$ space reveals two overlapping spectra corresponding to subluminal and superluminal wave packets (compare to the displaced spectra for Ch1 and Ch2 in Fig.~\ref{Fig:Spectra}c). With free propagation, the two wave packets separate as in the cases of wave packets in different spectral channels (Fig.~\ref{Fig:PulsePropagation2} and Fig.~\ref{Fig:PulsePropagation3}a,b). Furthermore, the wave packets can be located in different \textit{spatial} channels (along the transverse $x$-coordinate) by structuring the SLM spectral phase distribution along the direction that encodes the spatial frequencies $k_{x}$. An example is shown in Fig.~\ref{Fig:PulsePropagation3}d where two ST wave packets are synthesized with group velocities $\widetilde{v}_{1}$ and $\widetilde{v}_{2}$ but with SLM phase distributions that are displaced vertically (rather than interleaved as in Fig.~\ref{Fig:PulsePropagation3}c), resulting in two wave packets that are in turn displaced along the transverse $x$-axis by $\sim\!1.4$~mm.

A unique aspect of our strategy -- in comparison with slow-light and fast-light approaches -- is the absence of fundamental bandwidth constraints. Because we do not exploit an optical resonance, the bandwidth is in principle limited only by the experimental resources available. Increasing the utilized bandwidth (reducing the pulse width) does not affect the maximum differential group delay that is determined primarily by the spectral uncertainty \cite{Yessenov19OE}. To verify this, we confirm that the same group delay is obtained for a ST wave packet of fixed $\widetilde{v}$ as the pulse width is reduced from $\sim\!4$~ps to $\sim\!1$~ps, corresponding to an increase in the DBP by a factor of $\sim\!4$. Here the phase distribution imparted to Ch1 is extended across all three spectral channels to a $\sim\!1.5$-nm bandwidth.

\begin{figure*}[t!]
  \begin{center}
  \includegraphics[width=16.6cm]{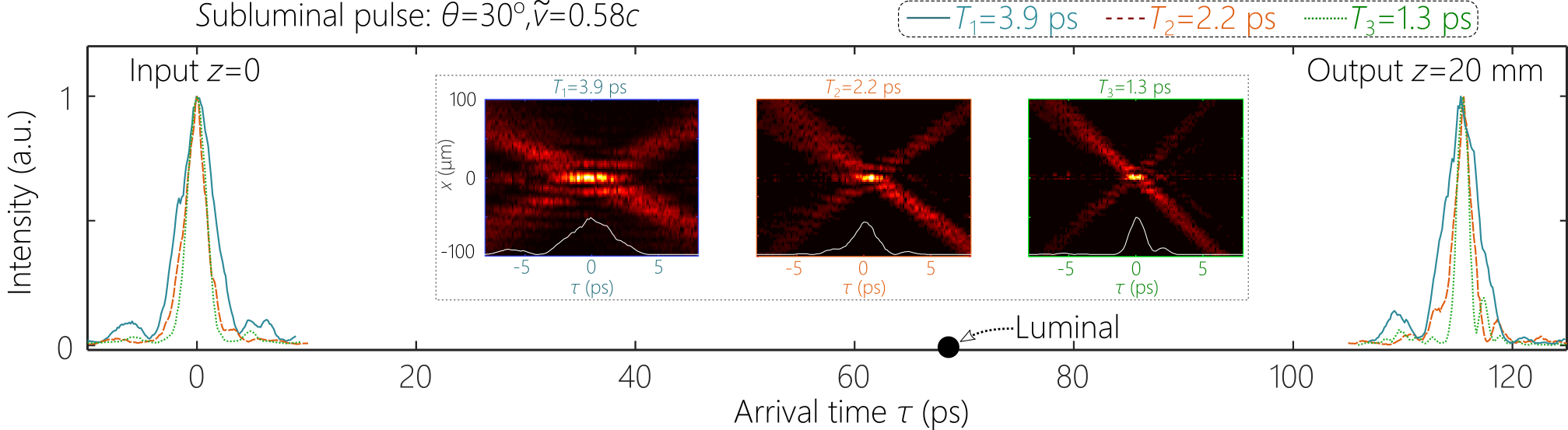}
  \end{center}
  \caption{Increase in the DBP with reduced pulse width. We plot the on-axis intensity profile $I(0,z,\tau)$ of three ST wave packets of different widths ($T_{1}\!\approx\!3.9$~ps, $T_{2}\!\approx\!2.2$~ps, and $T_{3}\!\approx\!1.3$~ps) but identical group velocity ($\widetilde{v}\!=\!0.58c$) recorded at two detector positions: $z\!=\!0$ (left) and $z\!=\!20$~mm (right). The pulses undergo the same delay of $\sim\!115$~ps with minimal pulse reshaping. Insets are the spatio-temporal profiles of the wave packets.}
  \label{Fig:DBP}
\end{figure*}

The measurement results are shown in Fig.~\ref{Fig:DBP}. First, we synthesize three ST wave packets of on-axis temporal widths of $T_{1}\!\approx\!4$~ps, $T_{2}\!\approx\!2.2$~ps, and $T_{3}\!\approx\!1.3$~ps, corresponding to bandwidths of 0.5~nm (Ch1), 1~nm (Ch1+Ch2), and 1.5~nm (Ch1+Ch2+Ch3), respectively. The insets in Fig.~\ref{Fig:DBP} show the spatio-temporal profiles of the three wave packets, all having the same group velocity $\widetilde{v}\!=\!0.58c$ ($\theta\!=\!30^{\circ}$). Next we trace the on-axis intensity for two positions of the detectors: an initial position of $z\!=\!0$ and then at $z\!=\!20$~mm. The group delay experienced by the three ST wave packets $\sim\!115$~ps is independent of the pulse width, resulting in a linear increase in DBP with the bandwidth (inversely with the pulse width). Here, the values of the DBP are $\approx\!29$, 52, and 89 respectively.

\begin{figure}[t!]
  \begin{center}
  \includegraphics[width=8.6cm]{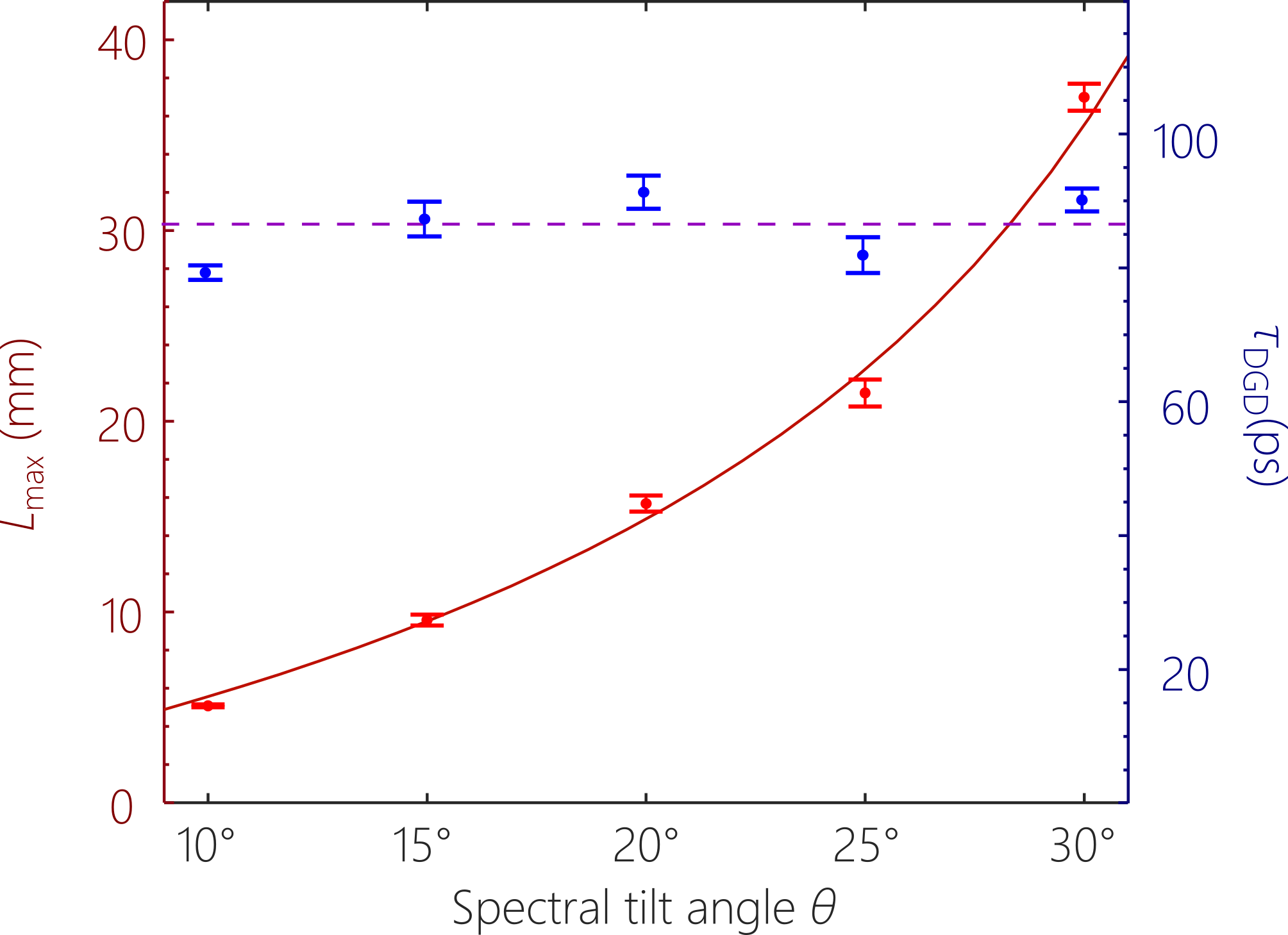}
  \end{center}
  \caption{Propagation distance $L_{\mathrm{max}}$ and differential group delay $\tau_{\mathrm{DGD}}$ against the spectral tilt angle $\theta$. The propagation distance is extracted from the on-axis time-averaged intensity profiles $I(x,z)$, where $I(0,L_{\mathrm{max}})\!=\!\tfrac{1}{e}I(0,0)$. The error bars correspond to $\delta L\!=\!\Delta z/\!\sqrt{2}$, where $\Delta z$ is the measurement step along $z$. }
  \label{Fig:Lmax}
\end{figure}

Finally, we examine the limit on miniaturizing the physical length of the delay line. By increasing the deviation of the group velocity $\widetilde{v}\!=\!c\tan{\theta}$ from $c$ (i.e., the deviation of $\theta$ from $45^{\circ}$), a large differential group delay can be realized over a short propagation distance. We plot in Fig.~\ref{Fig:Lmax} the measured propagation distance $L_{\mathrm{max}}$ and the associated differential group delay for subluminal pulses with respect to a luminal pulse. The measurements made use of ST wave packets with spectral tilt angles in the range $10^{\circ}\!<\!\theta\!<\!30^{\circ}$, corresponding to group velocities in the range $0.017c\!<\!\widetilde{v}\!<\!0.58c$. Note that the differential group delay remains fixed at $\sim\!85$~ps independently of $\theta$, $\widetilde{v}$, and the pulse width $T$, and is instead determined by the spectral uncertainty \cite{Yessenov19OE}. At the lower limit of $\theta\!=\!10^{\circ}$, the differential group delay is achieved over a distance of only 5~mm. This demonstrates the possibility of constructing compact optical delay lines with large DBP.

\section*{Discussion}

We have shown here that ST wave packets can provide a verstaile platform for large controllable inline differential group delays in free space by assigning different group velocities through spatio-temporal spectral-phase modulation. In this regard, ST wave packets offer the advantage of symmetrizing the treatment of the subluminal and superluminal regimes, which are equally useful in an optical buffer, in contrast to the divergence between the physical realizations of slow-light and fast-light (with the latter usually being significantly more challenging). The ST wave packets undergo minimal spreading in space and time, and hence minimal attenuation, while undergoing in one example a differential group delay of $\sim\!160$~ps between a pair of ST wave packets over a distance of 40~mm. To put this value in perspective, we note that spatial structuring in Refs.~\cite{Giovannini15Science,Bouchard16Optica} resulted in differential group delays of $\sim\!10$'s of femtosecond over a distance of 1~m, so that our result presents an improvement of $\sim\!6$ orders-of-magnitude; whereas in Refs.~\cite{Bonaretti09OE,Bowlan09OL,Kuntz09PRA}, X-waves revealed differential group delays of $\sim\!100$'s of femtoseconds over $\sim\!10$~cm, so that our result corresponds to a boost of $\sim\!4$ orders-of-magnitude.

Such an optical delay line is a critical component of an optical buffer. Several crucial steps remain before achieving this goal, including extending this experimental realization to the telecommunications band at $\sim\!1.5$~$\mu$m, making use of data streams to test the possibility of alleviating data packet contention at a switch, in addition to evaluating the required resources for our system (especially in terms of overall physical footprint) in comparison to alternative approaches. Furthermore, extending the transverse confinement of ST wave packets to both transverse directions (an optical needle) rather than the single dimension (light sheet) demonstrated here will enable their coupling to waveguides and other photonic devices.

The unique characteristics of ST wave packets demonstrated here pave the way to such exotic possibilities as colliding co-propagating or even counter-propagating optical pulses produced by the same source. Such scenarios could be useful in a variety of applications, including laser machining, laser fusion, and electron acceleration \cite{Jolly19OL}. Finally, we note that another promising approach for the spatio-temporal synthesis of optical wave packets of controllable group velocity has also been recently proposed and investigated \cite{SaintMarie17Optica}.

\bibliography{Main}

\vspace{2mm}
\noindent
\textbf{Acknowledgments}\\
We thank Matteo Clerici, Kenneth L. Schepler, and Abbas Shiri for helpful discussions. This work was supported by the U.S. Office of Naval Research (ONR) under contracts N00014-17-1-2458 and N00014-19-1-2192.

\vspace{2mm}
\noindent
\textbf{Author contributions}\\
\noindent
A.F.A. and M.Y. developed the concept and designed the experiment. M.Y. carried out the experiment and analyzed the data. B.B. assisted with the experiment. A.F.A. and P.J.D. supervised the research. All authors contributed to writing the paper.

\noindent
Correspondence and requests for materials should be addressed to A.F.A.

\noindent
(email: raddy@creol.ucf.edu).

\vspace{2mm}

\noindent
\textbf{Competing interests:} The authors declare no competing interests.

\newpage

\section*{Methods}

\subsection*{Theory for ST wave packets}

\begin{figure*}[t!]
  \begin{center}
  \includegraphics[width=17.6cm]{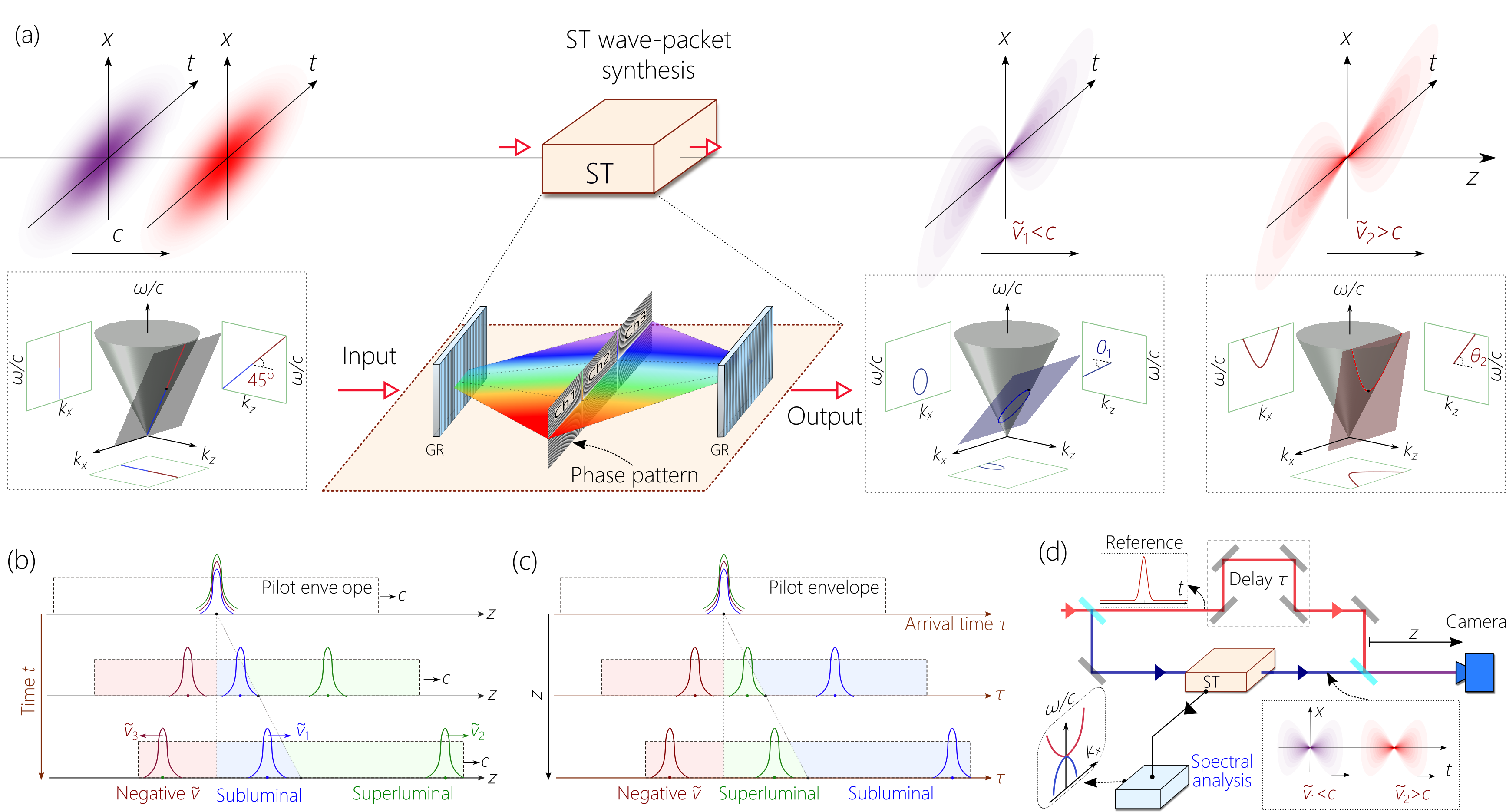}
  \end{center}
  \caption{Concept of a free-space optical delay line using ST wave packets. (a) Overall layout. Traditional pulses whose group velocities are equal ($\widetilde{v}\!=\!c$) are converted into ST wave packets each assigned a different group velocity (subluminal $\widetilde{v}_{1}\!<\!c$ and superluminal $\widetilde{v}_{2}\!>\!c$), so that they separate in time upon free propagation. The spatio-temporal spectrum of the traditional pulses and ST wave packet are depicted on the surface of the light-cone $k_{x}^{2}+k_{z}^{2}\!=\!(\tfrac{\omega}{c})^{2}$ and their projections are also shown (straight line for a plane-wave pulse, an ellipse for the subluminal wave packet, and a hyperbola for the superluminal wave packet). Here $\widetilde{v}\!=\!c\tan{\theta}$, where $\theta$ is the spectral tilt angle. Inset shows the structure of the ST wave packet synthesizer; GR: Diffraction grating. (b) Dynamics of ST wave packets under the pilot envelope. The locations of subluminal, superluminal, and negative-$\widetilde{v}$ wave packets are shown along the propagation axis $z$ at different instances in time. (c) The arrival times $\tau$ of ST wave packets are shown for different detector locations along $z$. (d) Schematic of the optical setup for synthesizing and characterizing ST wave packets.}
  \label{Fig:Concept}
\end{figure*}

\begin{figure}[t!]
  \begin{center}
  \includegraphics[width=8.6cm]{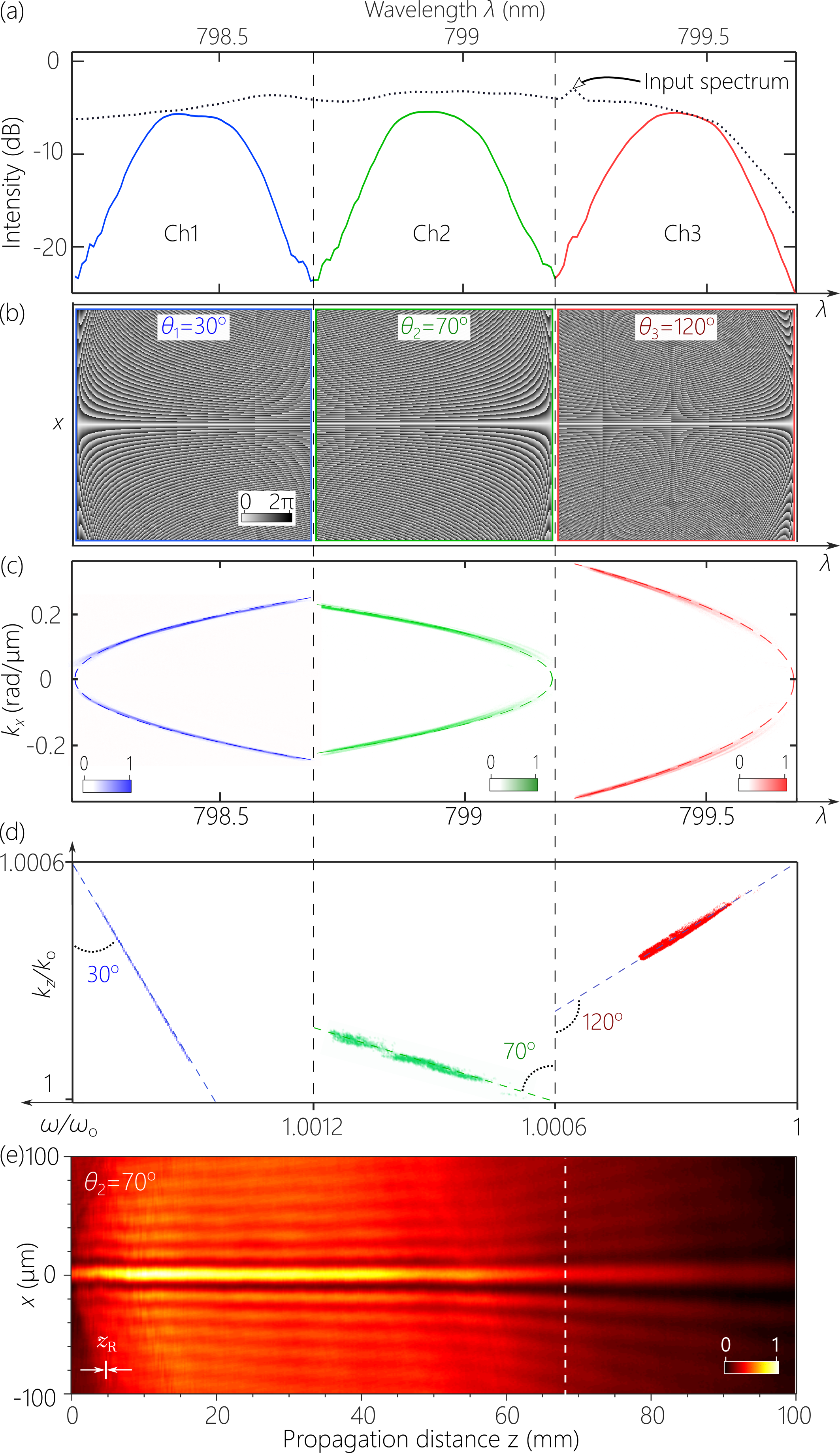}
  \end{center}
  \caption{(a) Spectra of the three spectral channels (solid curves) alongside the spectrum of the Ti:sapphire laser (dotted curve). (b) The 2D SLM phase distributions across the three channels to synthesize ST wave packets: subluminal, Ch1; superluminal, Ch2; and negative-$\widetilde{v}$, Ch3. (c) Measured spatio-temporal spectrum $|\widetilde{E}(k_{x},\lambda)|^{2}$ of the co-synthesized ST wave packets in the three spectral channels. Dashed curves are the theoretical expectations. (d) Spatio-temporal spectral projections onto the $(k_{z},\tfrac{\omega}{c})$-plane. The spectra for the different channels are shifted along the $k_{z}$-axis to fall within the same range for clarity. (e) Measured time-averaged axial evolution of the intensity $I(x,z)\!=\!\int\!dt|E(x,z,t)|^{2}$ of Ch2. Note the difference in scale along $x$ and $z$. The thin white bar in the bottom-left corner corresponds to the Rayleigh range $z_{\mathrm{R}}$ of a Gaussian beam having the same width as the ST wave packet. The vertical dashed line identifies $z\!=\!L_{\mathrm{max}}$.}
  \label{Fig:Spectra}
\end{figure}

\begin{figure*}[t!]
  \begin{center}
  \includegraphics[width=16.2cm]{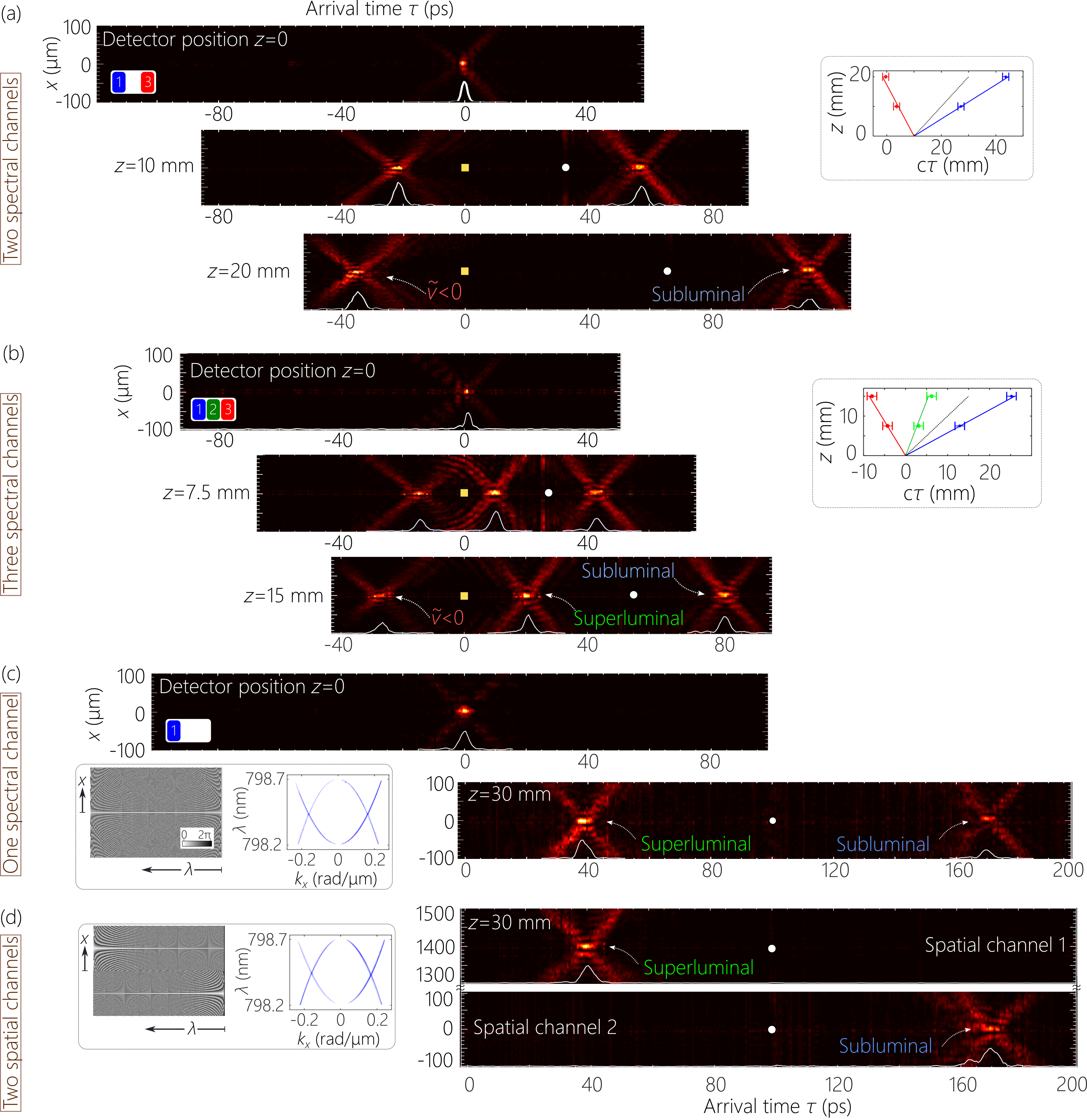}
  \end{center}
  \caption{\footnotesize{Propagation of ST wave packets in different or the same spectral or spatial channels. (a) The panels record the arrival time of subluminal (Ch1) and negative-$\widetilde{v}$ (Ch3) wave packets. The two wave packets arrive simultaneously with the detector placed at $z\!=\!0$. At $z\!=\!20$~mm, the subluminal wave packet arrives at $\tau\!\approx\! 113$~ps while the negative-$\widetilde{v}$ wave packet arrives at $\tau\!\approx\!-35$~ps (\textit{earlier} than the arrival time at the initial detector location of $z\!=\!0$). The white circle identifies the arrival time of a luminal pulse, whereas the yellow square identifies the $\tau\!=\!0$ position. Inset shows the dynamics of the wave-packet centers (similarly to Fig.~\ref{Fig:PulsePropagation2}b). (b) Same as (a) but with Ch1, Ch2, and Ch3 all activated. (c) Two wave packets (subluminal with $\theta_{1}\!=\!30^{\circ}$ and superluminal with $\theta_{2}\!=\!70^{\circ}$) occupy the \textit{same} spectral channel initially overlap when the detector is placed at $z\!=\!0$, and subsequently separate at $z\!=\!30$~mm. Inset shows the SLM 2D phase distribution required (interleaved distributions from Ch1 and Ch2 in Fig.~\ref{Fig:Spectra}b) and the measured overlapping spatio-temporal spectra. (d) Same as (c) but with the wave packets in two spatial channels displaced along $x$.}}
  \label{Fig:PulsePropagation3}
\end{figure*}

The spatio-temporal spectrum of a ST wave packet occupies a reduced-dimensionality spectral subspace with respect to traditional wave packets. We consider here one transverse dimension $x$, and the free-space dispersion relationship $k_{x}^{2}+k_{z}^{2}\!=\!(\tfrac{\omega}{c})^{2}$ is geometrically represented by the surface of the cone; here $z$ is the axial coordinate along the propagation direction, $\omega$ is the temporal frequency, and $c$ is the speed of light in vacuum. The spatio-temporal spectrum of a typical pulsed beam occupies a 2D patch on the surface of the light-cone corresponding to a finite spatial bandwidth (that determines the width of the beam profile) and a finite temporal bandwidth (that determines the pulse linewidth). A ST wave packet also has finite spatial and temporal bandwidths, but instead of being represented by a 2D patch on the light-cone, it lies along the intersection of the light-cone with a tilted spectral plane described by the equation:
\begin{equation}
\Omega\!=\!(k_{z}-k_{\mathrm{o}})\,c\,\tan{\theta},
\end{equation}
where $\Omega\!=\!\omega-\omega_{\mathrm{o}}$, $\omega_{\mathrm{o}}$ is a fixed frequency, and $k_{\mathrm{o}}\!=\!\omega_{\mathrm{o}}/c$ is the corresponding free-space wave number; see Fig.~\ref{Fig:Concept}a, right insets. As a result, each spatial frequency $k_{x}$ is associated with a single temporal frequency $\omega$, with a delta-function correlation whose functional form depends on $\theta$, $k_{x}\!=\!k_{x}(\Omega,\theta)$ \cite{Kondakci17NP,Yessenov19PRA}. Consequently, the envelope $\psi(x,z,t)$ of the wave packet $E(x,z,t)\!=\!e^{i(k_{\mathrm{o}}z-\omega_{\mathrm{o}}t)}\psi(x,z,t)$ becomes
\begin{equation}
\psi(x,z,t)=\int\!d \Omega\widetilde{\psi}(\Omega)e^{ik_{x}(\Omega,\theta)x}e^{-i\Omega(t-z/\widetilde{v})}=\psi(x,0,t-z/\widetilde{v}),
\end{equation}
which is a propagation-invariant wave packet traveling at a group velocity of $\widetilde{v}\!=\!c\tan{\theta}$, $\widetilde{\psi}(\Omega)$ is the Fourier transform of $\psi(0,0,t)$, and $k_{x}(\Omega,\theta)$ is the equation of the conic section at the intersection of the light-cone with the spectral plane after being projected onto the $(k_{x},\tfrac{\omega}{c})$-plane \cite{Yessenov19PRA}; see Fig.~\ref{Fig:Concept}a and Fig.~\ref{Fig:Spectra}c. The kind of conic section involved depends on the value of $\theta$: when $\theta\!=\!0$ it is a circle; when $0\!<\!\theta\!<\!45^{\circ}$ or $135^{\circ}\!<\!\theta\!<\!180^{\circ}$ an ellipse; when $45^{\circ}\!<\theta\!<\!135^{\circ}$ a hyperbola; when $\theta\!=\!45^{\circ}$ a straight line; and when $\theta\!=\!135^{\circ}$ a parabola. In all cases, the projection of the spatio-temporal spectrum onto the $(k_{z},\tfrac{\omega}{c})$-plane is a straight line making an angle $\theta$ with the $k_{z}$-axis; see Fig.~\ref{Fig:Spectra}d.

In our experiments we divide the available spectrum into three segments. Therefore, the synthesized ST wave packets are in general the sum of three individual ST wave packets having the form,
\begin{eqnarray}
\psi(x,z,t)&=&\psi_{1}(x,z,t)+\psi_{2}(x,z,t)+\psi_{3}(x,z,t)\nonumber\\
&=&\psi_{1}(x,0,t-z/\widetilde{v}_{1})+\psi_{2}(x,0,t-z/\widetilde{v}_{2})+\psi_{3}(x,0,t-z/\widetilde{v}_{3}).
\end{eqnarray}
Here $\widetilde{v}_{j}\!=\!c\tan{\theta_{j}}$, $j\!=\!1,2,3$, is the group velocity of each wave packet, and $\theta_{j}$ is the spectral tilt angle of the spectral plane associated with each wave packet. The envelope of each wave packet is given by
\begin{equation}
\psi_{j}(x,z,t)=\int\!d \Omega\widetilde{\psi}_{j}(\Omega)e^{ik_{x}(\Omega,\theta_{j})x}e^{-i\Omega(t-z/\widetilde{v})}=\psi_{j}(x,0,t-z/\widetilde{v}_{j}),
\end{equation}
where $\widetilde{\psi}_{j}(\Omega)$ is the Fourier transform of $\psi_{j}(0,0,t)$, and each spectral channel determines the domain of integration for $\Omega$.

In our experiments, we have $\theta_{1}\!=\!30^{\circ}$ (subluminal, conic section is an ellipse); $\theta_{2}\!=\!70^{\circ}$ (superluminal, hyperbola); and $\theta_{3}\!=\!120^{\circ}$. The spectrum that is spatially resolved by the diffraction grating and incident on the SLM has a bandwidth of $\sim\!1.5$~nm. The SLM active area is divided into three sections, each corresponding to $\sim\!0.5$~nm of spectrum (Fig.~\ref{Fig:Spectra}a). Within each one of these sections, a 2D phase pattern $\Phi_{j}$ is implemented (Fig.~\ref{Fig:Spectra}b) that produces the targeted spatio-temporal spectral association $k_{x}^{(j)}(\Omega,\theta_{j})$ (Fig.~\ref{Fig:Spectra}c).

\clearpage

\clearpage

\clearpage

\clearpage

\clearpage

\end{document}